\documentclass[floatfix,prb,aps,twocolumn,showpacs,preprintnumbers,amsmath,amssymb]{revtex4}  
\usepackage{graphicx}
\usepackage{dcolumn} 
\usepackage{exscale} 
\usepackage{bbm}  
\usepackage{amsfonts} 
 
\def\shiftdown#1{#1\llap{\lower.04ex\hbox{#1}}}

\begin{document} 
\noindent
{\large
\leftline{\bf Coherent feedback from dissipation:}
\leftline{\bf the lasing mode volume of random lasers}
}

\vspace*{-0.1cm}
\noindent
Regine Frank$^{1*}$, Andreas Lubatsch$^{2}$, Johann Kroha $^{2}$

{\small
\noindent
$^1$Institut~f\"ur~Theoretische Festk\"orperphysik,\\ 
\phantom{$^1$}KIT Karlsruhe, 76128 Karlsruhe, Germany\\
$*$email:regine.frank@kit.edu\\
$^2$Physikalisches Institut, Universit\"at Bonn,\\ 
\phantom{$^1$}Nussallee 12, 53115 Bonn, Germany\\
\noindent
}

\vspace*{0.1cm}
\noindent
{\bf In any quantum or wave system dissipation leads to
  decoherence. Therefore, it was surprising in first instance when experiments
  on strongly lossy random lasers showed unambiguously by measurements of the photon
  statistics and of the lasing mode volume that coherent feedback is possible
  in such systems \cite{Cao98,Cao99,Cao00}. In coherent-feedback lasers the photons
form a far-from-equilibrium condensate in the sense that a single quantum
state is occupied by a macroscopic number of photons \cite{John04}. We
demonstrate that the lossy dynamics of random lasers alone imply a finite
lasing mode volume, thus resolving the puzzle about coherent
feedback without resonator. Our theory of random lasing including nonlinear
gain and gain saturation predicts a characteristic dependence of this lasing mode volume on the pump intensity, which can be tested experimentally.
}
The concept of a random laser (RL), i.e., lasing in a homogeneously disordered,
laseractive medium without cavity, was introduced early on by 
Letokhov. It was then foreseen that light would be amplified 
by stimulated emission while diffusing through the random medium and 
covering an unlimited region in space. Consequently, a continuum of all the
modes whose frequency lies within the laser transition line would participate in
the lasing, leading to an uncorrelated (Bose-Einstein) photon number 
statistics (distributed feedback) and homogeneous laser emission from the
entire system volume. By contrast, it was found in experiments on ZnO 
powder lasers that the laser light emission occurs from spatially strongly 
confined lasing spots with sharp, discrete emission lines and Poissonian
photon number statistics, all three features being unambiguous signatures 
of coherent feedback, i.e. of laser emission from a single, spatially 
confined mode \cite{Kalt09}. This spatial confinement is at the heart of
understanding coherent feedback and the formation of a non-equilibrium 
photon condensate state in homogeneous random lasers. 
However, its origin remained obscure ever since its
discovery, because the commonly known, possible confinement mechanisms cannot 
explain the observed experimental facts. 
Anderson localization (AL) of light in disordered media can be ruled out,
because the photonic scattering mean free path $\ell$ is much longer than
the wavelength of light $\lambda$, so that AL does not occur in the systems
at hand, and the Ioffe Regel criterion is not met, i.e. $kl>1$. Random microcavities, preformed accidentially in the disordered 
medium, are unlikely to be the origin of coherent feedback, because 
neither would their intrinsic surface roughness allow for localized cavity 
modes to exist nor could they explain the observed dependence of the average 
lasing spot radius on the pump intensity \cite{Cao98,Cao99,Cao00}.

Here we present a genuinely new mechanism for generation of a finite
coherence or mode volume and coherent feedback, based on dissipation at the 
system surface. We show that optical losses at the surface imply that the
light field in the bulk of the system can be coherent (correlated) only over 
a finite length $\xi$. Hence, this defines a volume within coherent 
feedback occurs, leading to Poissonian photon statistics. We   
identify $\xi$ with the average radius of a lasing mode. It is dynamically 
generated by the highly non-linear, lossy dynamics of open laser systems
and, as such, exists only in the lasing state. As we perceived, this
confinement mechanism is closely related to the causality of both the propagation
of light and the transport of light intensity. Moreover, this being a dynamical effect, we 
predict $\xi$ to have a characteristic, namely power law, dependence on the 
laser pump intensity, which is found to be in qualitative agreement with 
available experiments\cite{Cao98,Cao99,Cao00}. 

\begin{figure}
  \begin{center}
    \includegraphics[width = \linewidth,clip]{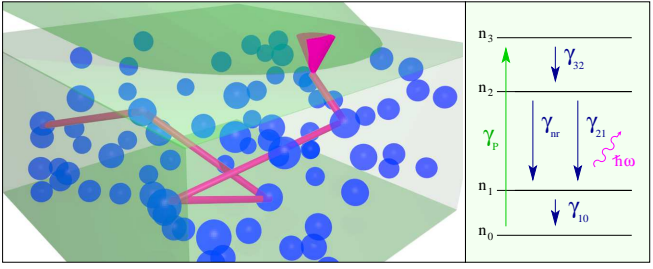}
    \caption{text
      {\bf a} Random Laser setup. The pump diameter is assumed to be large
      compared to the measured emission.
      {\bf b} Four level laser sceme.
    \label{fig1}
    }
  \end{center}
  \vspace*{-0.5cm}
\end{figure}

As a RL system we consider specifically a layer of compressed powder of 
laser-active material, whereby the grains act as random scatterers and at the
same time as amplifying medium. The layer has a thickness $d$ and 
extends infinitely in the x-y plane, as depicted in Fig.~\ref{fig1}a. 
The pump light covers a wide surface area and affects the entire layer,
so that the pump intensity may be assumed homogeneous 
across the entire volume of the considered system.  
The laser material is characterized by an atomic four-level scheme with transition 
rates $\gamma_{ij}$ as defined in
Fig.~\ref{fig1}b, although any other laser type would be possible. 

In order to address the problem of the coherence volume in random lasers
it is essential to set up the equation of motion for the coherent part of 
the radiation. This part is gerenerated by stimulated emission only,
although the quantum dynamics of the gain medium includes also spontaneous
emission. 
The time evolution of the electric radiation field can be parameterized by
as ${\bf E}(t,{\bf r}) = {\bf E}_0(t,{\bf r})\ exp[-i\omega t]$, where 
$\omega/2\pi$ is the light frequency and ${\bf E}_0(t,{\bf r})exp[-i\omega t]$ 
an amplitude function varying slowly on the scale of $2\pi/\omega$. 
We now observe that random lasers have to be treated as open systems, i.e.,
the loss-induced  damping time is always much shorter than the lifetime 
$\gamma_{12}$ of the upper atomic laser level (class B laser). In this 
strongly damped regime, the stimulated part of the polarization follows the
electric field instantaneously. Hence, the wave equation for the stimulated 
emission is local in time and can be written in terms of the 
dielectric function $\varepsilon ({\bf r}, {\bf E}_0(t,{\bf r}))$ as
\begin{equation*} 
[\frac{\varepsilon({\bf r}, {\bf E}_0(t,{\bf r}))}{c^2}\omega^2 + 
\nabla ^2 ] {\bf E}(t,{\bf r}) =0.
\end{equation*} 
Note that $\varepsilon ({\bf r}, {\bf E}_0(t,{\bf r}))$ incorporates the full, 
nonlinear laser dynamics through its dependence on the field amplitude 
${\bf E}_0(t,{\bf r})$ and depends on position ${\bf r}$ both explicitly
because of the random position of the dielectric scatterers and implicitly
through its dependence on ${\bf E}_0$. The lasing state is characterized 
by a negative imaginary part of $\varepsilon({\bf r},{\bf E}_0(t,{\bf
  r}))$. 
It is a priori not known and therefore determined by solving a four
level laser rate equation system (Siegmann) in steady state (see supplementary
informations). 

\begin{eqnarray*} 
\frac{\partial N_3}{\partial t} &=& \frac{N_0}{\tau_{P}}  - \frac{ N_3}{\tau_{32}}\\
\frac{\partial N_2}{\partial t}&=& \frac{N_3}{\tau_{32}} - \left(\frac{1}{\tau_{21}} + \frac{1}{\tau_{nr}}\right)N_2 - \frac{\left( N_2 -N_1\right)}{\tau_{21}} n_{ph} \\
\frac{\partial N_1}{\partial t}&=& \left(\frac{1}{\tau_{21}}+ \frac{1}{\tau_{nr}}\right)N_2  + \frac{\left( N_2 -N_1\right)}{\tau_{21}} n_{ph} - \frac{ N_1}{\tau_{10}} \\
\frac{\partial N_0}{\partial t}&=&  \frac{N_1}{\tau_{10}} - \frac{N_0}{\tau_{P}} \\
N_{tot}&=& N_0 + N_1 + N_2 + N_3.
\end{eqnarray*}

The electronic transition
rates from the ground state to the uppermost state and from the lower lasing
level to the ground state are assumed to be large compared to all other
transitions. Thus we focus on the description of the photon number $n_{ph}$
and the inversion $n_2=N_2/N_{tot}=\frac{\gamma_P}{\gamma_P + \gamma_{nr} + \gamma_{21}\left(n_{ph}
  +1\right)}$ according to the Einstein rate equations in steady
state. $\gamma_P$ equals the pump rate, $\gamma_{nr}$ represents the
nonradiative decays and $\gamma_{21}= 1/\tau_{21}$ features the stimulated
emission rate, which is the inverse of the lifetime or relaxation time of the
upper lasing level. The rsulting inversion of the occupation number thus
defines the microscopic optical gain and the challenge is now to relate this
gain to the coherent (correlated) part of the diffusing radiation.
Therefor the  laser rate system is coupled to sophisticated model for microscopic
transport of light in disordered amplifying random media
\cite{Frank11,Frank06, Frank09, Lubatsch05} based
on Vollhardt-W\"olfle theory for the transport of electrons \cite{Woelfle80}, which allows for the consideration of the influences of the
microstructure of the amlifying media. For the completeness of the description of ramdom
lasing it has been demanded earlier by Florescu
et. al. \cite{John04} that the effects due to multiple scattering of light
and the internal Mie-resonances during these scattering events must be
considered. We included those contributions by generalizing the diagrammatic
description for the transport of light and we particularly focussed on
the influence of interferences. Therefore we had to deal with a Dyson
description of the propagation of the electromagnetic wave on the one hand and
the Bethe-Salpeter-Equation for the propagation and dissipation of the energy
density of light on the other.
The complete description can be found in the supplementary, but to introduce
the reader breefly to this fascinating subject, the Bethe-Salpeter-Equation
for the two particle Green's function or in other
words the energy propagator $\hat{\bf \Phi}$ is
given here - which looks a bit denaunting at first glance but actually it is a
treasure chest for the random laser theory.

\begin{eqnarray*}
\hat{\bf \Phi}
=
\left(\hat{\bf G}^R\otimes \hat{\bf G}^A\right)
\left[
\mathbbm{1}\otimes\mathbbm{1}
+
\hat{\bf \gamma}\hat{\bf \Phi}
\right].
\label{BS_Operator}
\end{eqnarray*}

The Bethe-Salpeter-Equation is the approbiate equation of motion for the
coherent part of light intensity $\Phi$. 
The Green's function ${\bf G}$ itself describes the behavior of the elctromagnetic
wave, whereas the so called irreducible vertrex $\hat{\bf \gamma}$ reveals all
interactions and the selfconsistence-loop is closed by the selfdependency of
$\hat{\bf \Phi}$. It has to be stressed that the propagation of light
intensity in diffusive or strong scattering regimes obeys the second law of
thermodynamics and thus the time scales of wave propagation and intensity
propagation in fact can be separated.
The results drawn from these considerations turned out to be really
promising, because the diffusive processes, the interferences and localization processes can be clearly
identified as separate contributions in the thereby derived diffusion constant
$D(\Omega)$

\begin{eqnarray*}  
D(\Omega)  
\left[1 - i \, \Omega  {\rm Re}\epsilon \omega \tau_a^2   \right] =   
D_0^{tot} - \tau_a^2 D(\Omega ) M(\omega ). 
\end{eqnarray*}

The first term on the right hand side $D_0^{tot}=D_0+D_b+D_s$ could be
slightly overlooked on the search for interference
effects which are represented through $M(\omega)$, the so called memory
kernel. 
A detailed study reveals, however, that the memory kernel cannot cause true
localization, i.e. $D=0$, in laser active media, and $D_0^{tot}$ cannot be compensated. 
$D_0$ reflects the bare diffusion, the
dissipative renormalizations due to absorbtion and/gain in the system are
summarized in the term $D_b$ and $D_s$. After all we remark that this diffusion constant
for systems, which are in any way laser-active, will never vanish, and therefore AL of
light in such systems seems to be beyond the question in the discussion of
random lasing.

From the microscopically derived correlation length $\xi$ (see supplementary
information) in steady state for
strong pumping (see
Fig. 4) one can draw several significant conclusions for the characteristics
of a random laser

\begin{eqnarray*}
\label{xi}
\xi = \frac{\alpha}{\sqrt{P}} + \xi_{\infty}
\end{eqnarray*}

where $\alpha$ is a numerical constant an $P$ the pumping.
It is found that the natural antagonists, the pump strength
$P$ and the loss at the surfaces play a severe though in first instance
counterintuitive role. The coherence length $\xi$ in steady state clearly
shrinks with the increase of the pump strength $P$ whereas dissipation or loss at the surfaces
of the depicted slab geometry (see Fig. 1) in contrary increases the
correlation length. 
The meaning of both results become obvious when we go back
to the roots of the idea of random lasing again. Dicke in 1968
proposed the possibility of the photon bomb by means of an infinite increase of the {$\bf
k$}-modes in an amplifying media. 
Our results in contrast proove that the
nature of the onset of lasing in amplifying random media is clearly described
by a finite correlation length and thus a finite mode volume which marks the
exponential decay of the spatially and spectrally coherent light intensity
inside the ZnO powder slab (Fig 3b).
Loss at the surface however leads to an increase of the correlation length
$\xi$. Both aspects, the dependency on the pumping and on the loss lead to gain
saturation which govern the correlation volume in steady state.

These results clearly forbid the development of a photon bomb in lossy, finite, but
amplifying random systems on the one hand, but also guarantee the coherence of
the emission. The relation between the atomic population inversion and the
dynamics of the coherent part of the light field described by the
Bethe-Salpeter-Equation constitutes the parameter-free link between microscopic and
mesoscopic processes in random lasing which has been missing up to the present. The
description of growth of the photon number density in both approaches establishes the selfconsistency chain
that incorporates the full dynamics of the random laser. Hence the first
parameter-free description of random lasing is established.

{\bf Acknowledgments.}
This work was supported by the Deutsche Forschungsgemeinschaft through
FOR~557 and grant no. KR1726/3-1,-2.
\noindent
{\bf METHODS}
 
\newpage
\begin{figure}
  \begin{center}
    \includegraphics[width = \linewidth,clip]{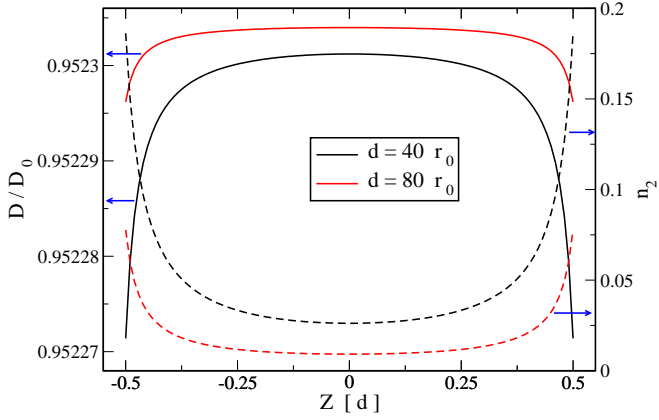}
    \caption{
      The full and self-consistent diffusion constant $D$ in units of the bare diffusion
 constant $D_0$
 is shown as a function of the width $Z$ across the sample
 for pump rate $P / \gamma_{21}= 2  $. 
 The sample width is
 set to be $d = 40 r_0$ and  $d = 80 r_0$ respectively as indicated in
 the legend , where $r_0$ is the radius of the scatterers. 
 As discussed in the text, the ratio 
 $D / D_0$ remains practically unchanged, since the relative change observed in the graph is
 of the same order as the accuracy, involved in the numerical evaluation.
 This implies that interference effects do not change as a function of the
 sample width.The diffusion constant $D$ can be seen to behave inversed compared
 to the inversion $n_2$. Both quantities show a weak dependency to the sample's
 width.
    \label{fig2}
    }
  \end{center}
  \vspace*{-0.5cm}
\end{figure}
\begin{figure}
  \begin{center}
    \includegraphics[width = \linewidth,clip]{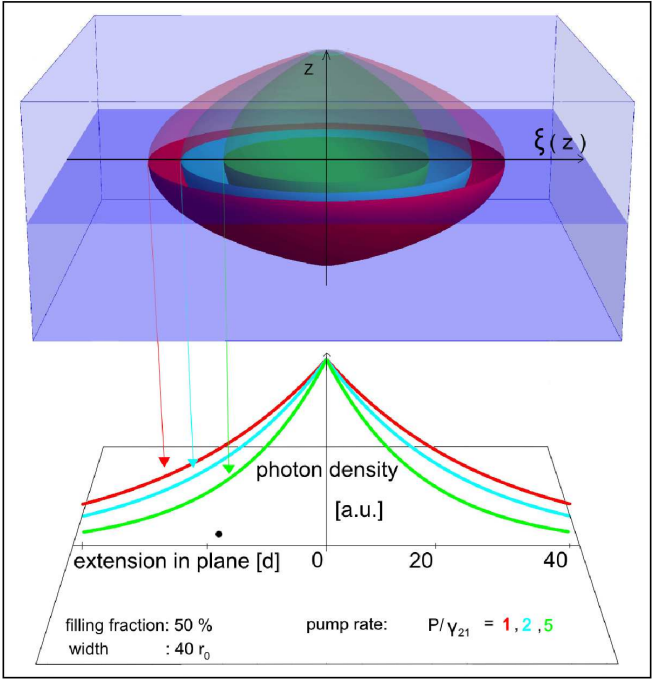}
    \caption{
      {\bf a} The correlation volume of the RL mode strongly decays with
      encreasing pump strength.  
      {\bf b} Calculated photon density 
    \label{fig3}
    }
  \end{center}
  \vspace*{-0.5cm}
\end{figure}
\begin{figure}
  \begin{center}
    \includegraphics[width = \linewidth,clip]{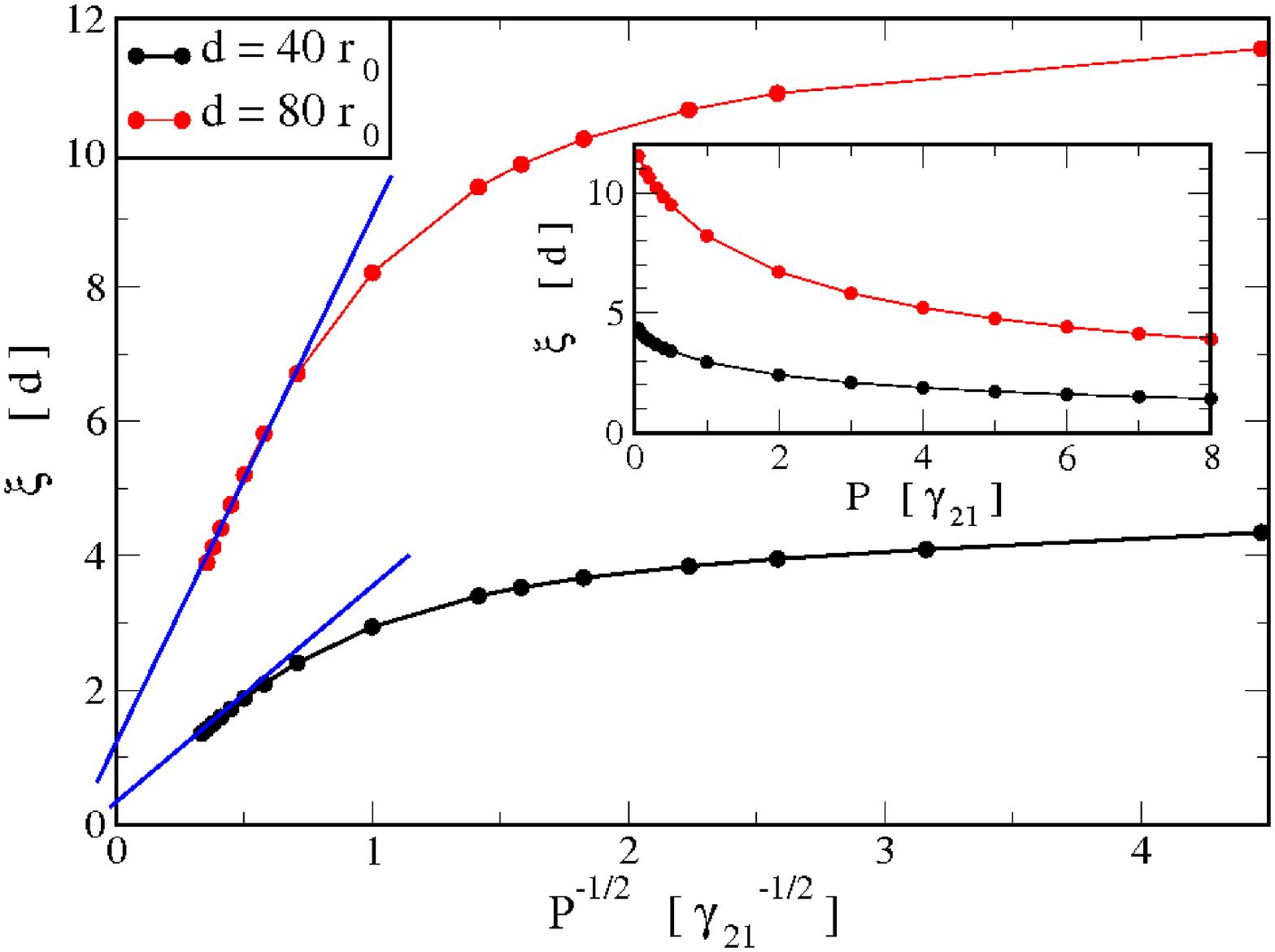}
    \caption{
       Correlation length $\xi$  as a function of 
 the inverse square root of the  pump rate, measured in units  of the inverse square 
root of the transition rate $\gamma_{21}$,
 at the sample surface, i.e. $Z=0.5d$.\newline
 The blue lines serve as a guide to the eye, emphasizing the linear behavior of
 the correlation length in this plot. This clearly reveals an inverse square
 root behavior of  $\xi$  as a function of pumping above threshold.
    \label{fig4}
    }
  \end{center}
\vspace*{-0.5cm}
\end{figure}
\noindent

\end{document}